\documentclass[aps,pra,onecolumn,superscriptaddress,nofootinbib]{revtex4-2}

\usepackage{amsmath,amssymb,graphicx, physics, comment, color}
\usepackage{units}
\usepackage[hidelinks]{hyperref}
\usepackage{upgreek}

\begin{document}

%\title{473 nm-Pumped Source of 680/1550 nm Entangled Photons for Fiber-Based Quantum Key Distribution}

\title{Highly Nondegenerate Entangled Photon Source for Fiber-Based Quantum Key Distribution}

%======
\author{Vasile-Laurențiu~Dosan}
\email{l.dosan@qo-jena.com}
\affiliation{Quantum Optics Jena GmbH, Am Zementwerk~8, 07745 Jena, Germany}

\affiliation{Institute of Applied Physics, Friedrich Schiller University Jena, Albert-Einstein-Str.~6, 07745 Jena, Germany}

\affiliation{Max Planck School of Photonics, Friedrich Schiller University Jena, Albert-Einstein-Str.~15, 07745 Jena, Germany}

%======
\author{Alek~Lagarrigue}
\affiliation{Quantum Optics Jena GmbH, Am Zementwerk~8, 07745 Jena, Germany}
%======
\author{Alessandro~Zannotti}
\affiliation{Quantum Optics Jena GmbH, Am Zementwerk~8, 07745 Jena, Germany}

%======
\author{Tushar~Parab}
\affiliation{Quantum Optics Jena GmbH, Am Zementwerk~8, 07745 Jena, Germany}

%======
\author{Yannick~Folwill}
\affiliation{Quantum Optics Jena GmbH, Am Zementwerk~8, 07745 Jena, Germany}

%======
\author{Fabian~Steinlechner}

\affiliation{Institute of Applied Physics, Friedrich Schiller University Jena, Albert-Einstein-Str.~6, 07745 Jena, Germany}

\affiliation{Max Planck School of Photonics, Friedrich Schiller University Jena, Albert-Einstein-Str.~15, 07745 Jena, Germany}

\affiliation{Fraunhofer Institute for Applied Optics and Precision Engineering, Albert-Einstein-Str.~7, 07745 Jena, Germany}

%======
\author{Oliver~de~Vries}
\affiliation{Quantum Optics Jena GmbH, Am Zementwerk~8, 07745 Jena, Germany}

\begin{abstract}
\noindent Entangled photon sources (EPSs) are essential building blocks for scalable quantum communication and quantum key distribution (QKD). We present a stable, highly nondegenerate EPS based on type-0 spontaneous parametric down-conversion (SPDC) in a crossed-crystal configuration, generating photon pairs at 680~nm and 1550~nm when pumped by a 473~nm laser. This wavelength combination, reported here for the first time, simultaneously benefits from the peak detection efficiency of the most of the Si-SPADs in the visible/near-infrared spectral range and the low-loss fiber transmission of the telecom C-band. This configuration provides the most favorable balance between performance and cost for detection using Si-SPADs and InGaAs detectors. The source exhibits a measured spectral bandwidth of 300~GHz, corresponding to a spectral brightness of up to $1.9\times 10^3$~pairs~s$^{-1}$~mW$^{-1}$~GHz$^{-1}$. Heralding efficiencies reach 18~\% (signal) and 34~\% (idler) with Si-SPAD and superconducting nanowire single-photon detectors (SNSPD) detection. The entangled state achieves visibilities of $(97.3\pm 1.0)\,\%$ in the H/V basis and $(94.9\pm1.6)\,\%$ in the D/A basis, yielding a fidelity of $\geq(96.1\pm1.3)\,\%$. These results establish the presented EPS as a practical wavelength-hybrid platform for fiber-based QKD and emerging long-haul quantum network architectures.

\end{abstract}

\maketitle

\section{Introduction}

Entanglement serves as a fundamental resource for quantum key distribution (QKD), enabling secure communication rooted in the principles of quantum nonlocality. In entanglement-based protocols such as E91 \cite{E91} or BBM92 \cite{bennett1992quantum}, pairs of entangled photons are distributed to two distant parties, who perform measurements in mutually unbiased bases. The strong correlations between their outcomes form the foundation for generating a shared secret key. As such, entanglement is a valuable physical resource for building scalable and provably secure quantum networks. Advances in high heralding entangled photon sources (EPSs) and long-distance distribution have made such protocols increasingly practical \cite{yin2020entanglement, valivarthi2016quantum, 404km, 421km}.

Spontaneous parametric down-conversion (SPDC) is a widely adopted method for generating entangled photon pairs, relying on three-wave mixing in nonlinear $\chi^{(2)}$ crystals. It generates states with strong correlations in polarization, energy, and momentum, making it ideal for quantum communication and QKD. SPDC remains the standard for high-quality EPSs due to its reliability and versatility in bulk crystal implementations~\cite{anwar2021entangled, Bilbao_paper}.

In fiber-based quantum communication, the 1550 nm (C-band) and 1310 nm (O-band) wavelengths are commonly used due to their compatibility with standard SMF-28 single-mode fibers. The C-band offers low attenuation (approximately 0.2 dB/km) but suffers from significant chromatic dispersion (around 17 ps/nm/km), which can degrade timing precision. In contrast, the O-band at 1310 nm exhibits near-zero dispersion, preserving temporal correlations, but has higher attenuation (about 0.35 dB/km). The choice between these bands involves a trade-off between transmission loss and dispersion effects~\cite{agrawal2012fiber}.

Single-photon detectors are a key component in quantum communication systems, with different technologies presenting trade-offs in performance, usability, and integration. For telecom wavelengths, superconducting nanowire single-photon detectors (SNSPDs) and InGaAs-based single-photon avalanche diodes (InGaAs-SPADs) are deployed. SNSPDs offer high detection efficiency $> 90\,\%$, short dead times $<25$ ns, and low dark counts $< 100$ Hz at 1550 nm, making them a good candidate for quantum applications. However, they require cryogenic cooling, have a large physical footprint, consume substantial power, and demand periodic maintenance, which limits their deployment in compact or commercial systems~\cite{natarajan2012snsps}. In contrast, InGaAs-SPADs operate at room or moderate cooling temperatures, are compact, and support continuous 24/7 operation. Yet, they typically exhibit lower detection efficiencies, higher dark count rates, and longer dead times ~\cite{hadfield2009single}. For the visible/NIR detection, Si-SPADs provide high efficiency, low timing jitter, and compact form factors, but are only suitable for visible and near-visible wavelengths, impending their use in telecom-band applications~\cite{cova2004history}.

A compelling strategy for practical QKD systems, that are industrial-grade, have a negligible down-time, and are cost-effective, is to use highly nondegenerate entangled photon pairs. In this approach, one photon lies within the Si-SPAD detection range (visible or NIR) and its partner in the telecom band. This enables the use of InGaAs-SPADs to detect the IR photon, while harnessing the high efficiency and low noise of Si-SPADs for the visible or NIR photon. A popular choice involves a 532 nm pump to generate photon pairs at 810 nm and 1550 nm via SPDC in bulk crystals such as ppLN or ppKTP~\cite{hentschel2010three, sauge2008single}. This configuration facilitates long-distance fiber transmission in the telecom C-band, while retaining efficient local detection in the NIR range.

In this work, we demonstrate a crossed-crystal EPS pumped at 473 nm, generating highly nondegenerate photon pairs at 680 nm and 1550 nm, a combination that, to the best of our knowledge, has not been reported in the literature. This configuration provides a compact and stable architecture, showing a stable operation with high entanglement fidelity ${\cal F} \geq 96\,\%$. As many Si-SPADs have a peak efficiency in the range of 550 to 700 nm, the presented EPS leverages the best efficiency point of the visible single photon detection. Moreover, it makes use of the low fiber attenuation at 1550 nm. The use of a diode-pumped solid-state laser (DPSSL) at 473 nm ensures single-line operation and spectral stability, suitable for long-term operation. The generated photon pairs exhibit narrow emission bandwidth of 300 GHz which enables high spectral brightness and reduced dispersion in long-distance fiber transmission. In the following, we describe the source design and performance model, the experimental setup, and the measured performance of the source.

\section{Source design and experimental implementation}
\label{sec:source_design}

\begin{figure}[h!]
\centering
\includegraphics[width=0.7\linewidth]{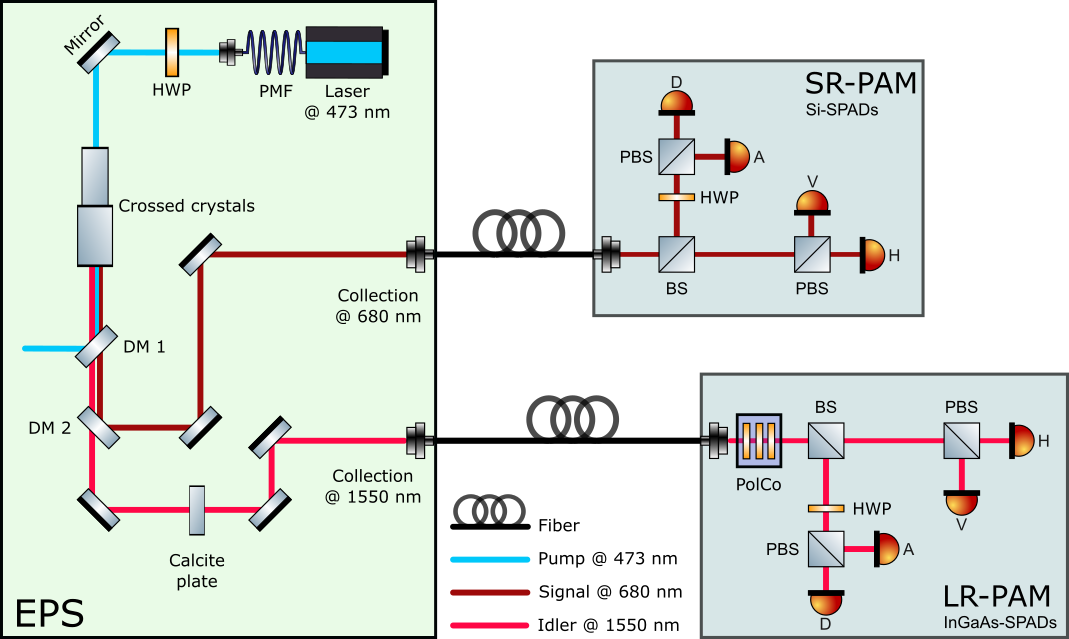}
\caption{\label{fig:experimental_setup} Setup for a wavelength-hybrid EPS: a 473 nm pump laser drives crossed nonlinear crystals to generate 680 nm/1550 nm entangled photon pairs, DM1 removes the pump, DM2 separates signal and idler which are analyzed via SR-/LR-PAM with Si- and InGaAs-SPAD detection. PMF: polarization-maintaining fiber, HWP: half-wave plate, DM: dichroic mirror, SR-/LR-PAM: short-range-/long-range-polarization analysis module, BS: beam splitter, PBS: polarization beam splitter, PolCo: polarization compensation.}
\end{figure}

The experimental setup is depicted in Fig.~\ref{fig:experimental_setup}. The EPS uses a fiber-coupled 473 nm DPSSL with a maximum output power of 40 mW. The laser features a narrow linewidth ($<$ 1 MHz), high spatial mode purity, minimal mode hopping, and excellent long-term power and wavelength stability, characteristics critical for phase-matched generation of biphoton pairs via SPDC.

The entangled photon source is based on a crossed-crystal configuration utilizing two periodically poled potassium titanyl phosphate (ppKTP) crystals, each with a length of 10\,mm and a cross-sectional aperture of $1 \times 1\,\mathrm{mm}^2$. The crystals are oriented with orthogonal optical axes to enable polarization entanglement through type-0 nondegenerate SPDC when pumped at 473 nm. Therefore, the output state becomes:
\begin{equation}
    \ket{\Psi} = \frac{1}{\sqrt{2}}\left(\ket{H_{680\,\mathrm{nm}} H_{1550\,\mathrm{nm}}} + e^{i\phi} \ket{V_{680\,\mathrm{nm}} V_{1550\,\mathrm{nm}}} \right),
\end{equation}
\noindent where \( H \) and \( V \) denote horizontal and vertical linear polarizations, respectively. The subscripts \( 680\,\mathrm{nm} \) and \( 1550\,\mathrm{nm} \) indicate the wavelengths of the signal and idler photons, respectively. The phase \( \phi \) is determined by optical path differences and birefringent effects within the system. It is homogenized over the emission spectrum by introducing a birefringent calcite plate with a length of 3.6\,mm in the idler arm. This approach compensates the temporal and phase walk-off between the two SPDC processes, following the method described in \cite{calcite}.

After the crystals, the residual pump beam is removed using a dichroic mirror (DM1) that reflects the pump while transmitting the down-converted photons. The signal and idler photons are subsequently separated using a second dichroic mirror (DM2), which reflects the signal and transmits the idler. After separation, both paths are coupled into single-mode fibers for the respective wavelength. 

To design the optimal focusing parameters, we followed the approach presented in \cite{Guerreiro_2013}. The focusing parameter $\xi$ is a dimensionless quantity that characterizes the spatial confinement of the pump beam with a wavelength of $\lambda_p$ and a waist of $w_p$, inside the nonlinear crystal of length $L$. It is defined as \cite{Fazili:24}:
\begin{equation}
    \xi = \frac{\lambda_p L}{2\pi w_p^2}.
\end{equation}
For the pump beam, we chose a focusing parameter of $\xi = 0.02$, which corresponds to a pump waist of $w_p = 194~\upmu\text{m}$ inside the nonlinear crystal. By projecting the biphoton state into a Gaussian mode and using simulated emission angles, we obtained optimal collection waists of $w_s = 87~\upmu\text{m}$ for the signal and $w_i = 141~\upmu\text{m}$ for the idler.

The polarization analysis modules (PAMs) are used to characterize the entangled state by measuring single-photon correlations in two mutually unbiased bases: horizontal/vertical (H/V) and diagonal/anti-diagonal (D/A). A beam splitter (BS) probabilistically routes photons to one of the two analysis paths. In the H/V path, a polarizing beam splitter (PBS) projects the state onto the H/V basis. In the D/A path, a half-wave plate (HWP) set at 22.5° rotates the polarization before the projection by a PBS. The short-range (SR) PAM uses Si-SPADs, while the long-range (LR) PAM utilizes InGaAs-SPADs. Active polarization compensation (PolCo), based on liquid crystals, is implemented in the LR-PAM to counteract polarization changes induced by the optical fiber. 

\section{Results}

\begin{figure}[h!]
\centering
\includegraphics[width=\linewidth]{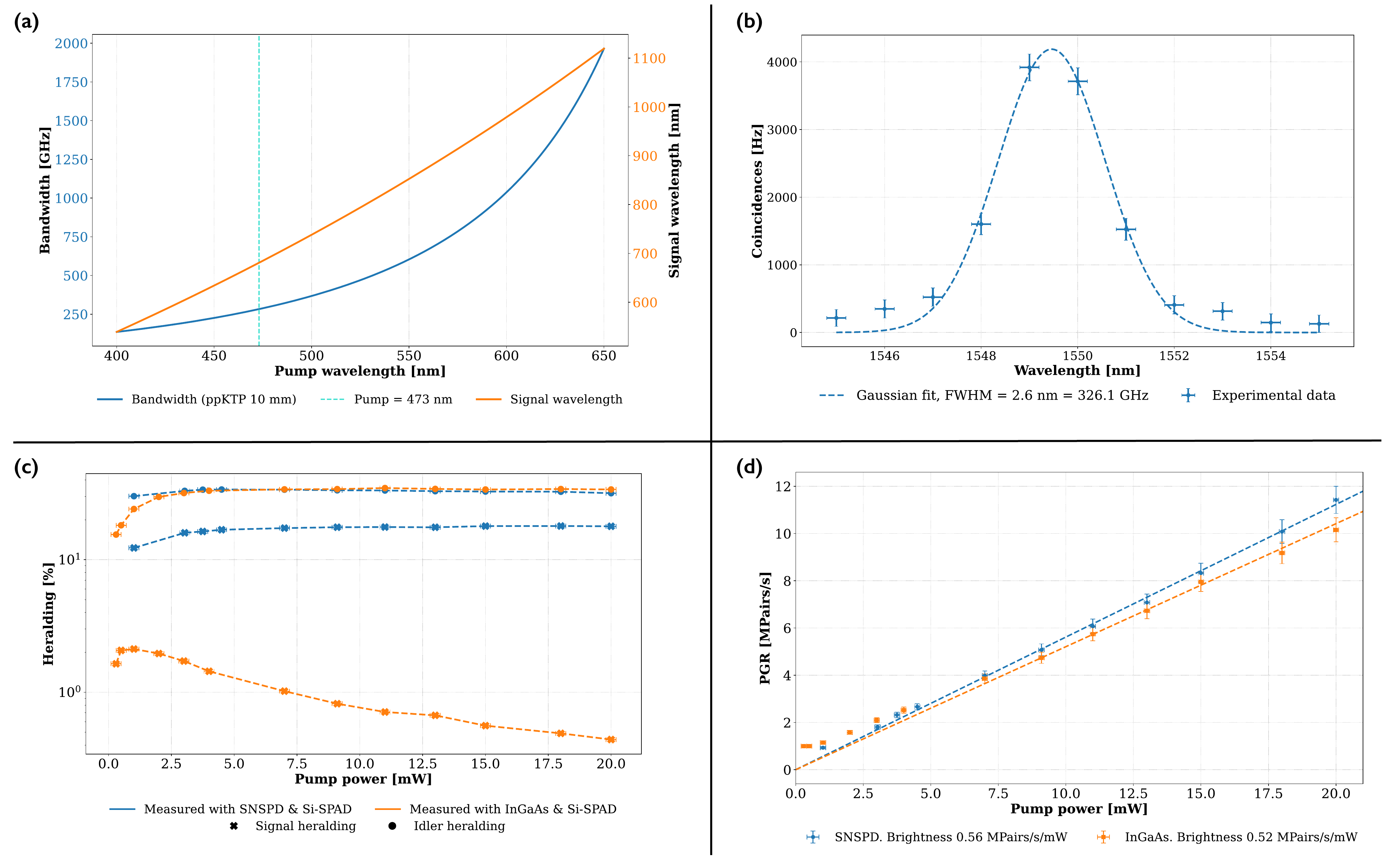}
\caption{\label{fig:results} (a) Emission bandwidth (left axis) and signal wavelength (right axis) vs pump wavelength for a 10 mm ppKTP crystal at 40 °C with fixed idler at 1550 nm. The vertical dashed line marks 473 nm pump. Bandwidth is derived from the FWHM of the SPDC phase-matching spectrum. (b) The measured idler spectra using a tunable filter with a bandwidth of 125 GHz. (c)~The signal and idler heralding as a function of pump power. (d) The photon generation rate (PGR) as a function of the pump power.}
\end{figure}

\subsection{Spectral bandwidth}

The spectral properties of SPDC are governed by the conservation of energy and momentum in a nonlinear optical medium \cite{Mandel_SPDC, anwar2021entangled, spdc_theory}. The spectral bandwidth arises from the finite length $L$ of the nonlinear crystal, which introduces a longitudinal phase-matching envelope with sinc\(^2\) dependence. It is defined as the full width at half maximum (FWHM) of the sinc$^2$ function which defines the emission spectra. The joint spectral intensity $I$ is given by:
\begin{equation}
I \propto \mathrm{sinc}^2\left( \frac{\Delta k \cdot L}{2} \right),
\end{equation}
\noindent where the phase mismatch is
\begin{equation}
\Delta k = k_{\text{pump}} - k_{\text{signal}} - k_{\text{idler}} - \frac{2\pi}{\Lambda},
\end{equation}
\noindent where $k_{\text{pump, signal, idler}}$ are the pump, signal, and idler wave vectors and $\Lambda$ is the poling period of the crystal.  

As shown in Fig.~\ref{fig:results}(a), the bandwidth becomes progressively narrower for shorter pump wavelengths. This trend reflects the stronger phase-matching constraints and steeper dispersion at shorter wavelengths. The calculations are performed for type-0 phase matching in a 10 mm ppKTP crystal, stabilised at \( 40^\circ \)C, with a fixed idler wavelength of 1550 nm. Furthermore, for a pump wavelength of 473 nm and an idler centered at 1550 nm, the corresponding signal wavelength is 680 nm, which is close to the peak detection efficiency of most Si-SPADs. Our simulations were performed using the wavelength- and temperature-dependent refractive indices described in \cite{Fan:87}.

The idler spectrum was characterized using a tunable optical bandpass filter with a 1~nm step size (Fig.~\ref{fig:results}(b)). The measured FWHM is 326 GHz (2.6~nm), slightly exceeding the 300~GHz theoretical bandwidth. This discrepancy arises from the finite spectral resolution of the filter, whose Gaussian-shaped passband has an intrinsic width of $\mathrm{FWHM}_{\mathrm{f}} \approx 125$~GHz. The measured spectrum is given by the convolution of the SPDC spectrum and the Gaussian spectral distribution of the filter. Here, we approximate both functions, with negligible errors, to be Gaussian and yield:
\begin{equation}
    \mathrm{FWHM}_{\mathrm{SPDC}} = \sqrt{\mathrm{FWHM}_{\mathrm{meas}}^{2} - \mathrm{FWHM}_{\mathrm{f}}^{2}},
\end{equation}
\noindent which gives a corrected idler bandwidth of 301 GHz (2.41 nm), in excellent agreement with the theoretical prediction (Fig.~\ref{fig:results}(a)). The central wavelength is 1549.5 nm, with temperature tuning enabling precise control of its spectral position.

\subsection{Heralding efficiency}

The coincidence rate between the signal and idler detection channels can be described as the sum of true and accidental coincidences. The true coincidence rate $ C_{\mathrm{true}}$, in the absence of timing constraints, is given by:
\begin{equation}
    C_{\mathrm{true}} = \text{PGR} \times \, \eta_{\mathrm{eff, s}} \, \eta_{\mathrm{eff,i}},
\end{equation}
\noindent where $\text{PGR}$ is the pair generation rate of the EPS, and $ \eta_{\mathrm{eff, s}}$ and $\eta_{\mathrm{eff, i}}$ are the effective heralding efficiencies for the signal and idler photons, respectively. These efficiencies account for all transmission losses, coupling losses, detector quantum efficiencies, and deadtime effects.

In practice, only a fraction of the true coincidences fall within the chosen coincidence time window $\Delta t$, due to the finite timing jitter $\sigma_{\mathrm{total}}$ of the detection system. Assuming Gaussian timing statistics, this fraction $F$ is \cite{detector_model_1}:
\begin{equation}
    F = \mathrm{erf} \left( \sqrt{\ln(2)} \, \frac{\Delta t}{\sigma_{\mathrm{total}}} \right),
\end{equation}
\noindent where $\mathrm{erf}(\cdot)$ denotes the error function.

Accidental coincidences arise when uncorrelated detection events occur within the same coincidence window. For low-probability detection per window ($S \, \Delta t \ll 1$), the accidental rate $C_{\mathrm{acc}}$ can be approximated by: 
\begin{equation}
    C_{\mathrm{acc}} \approx S_{\mathrm{measured,s}} \, S_{\mathrm{measured, i}} \, \Delta t,
\end{equation}
\noindent where $S_{\mathrm{measured,s}}$ and $S_{\mathrm{measured,i}}$ are the measured singles rates in the signal and idler channels, respectively, including noise contributions.

Finally, the measured coincidence rate $C_{\mathrm{measured}}$ is obtained by combining the fraction of captured true coincidences and the accidental coincidences:
\begin{equation}
    C_{\mathrm{measured}} = F \cdot C_{\mathrm{true}} + C_{\mathrm{acc}}.
\end{equation}
This model enables a direct comparison between theoretical expectations and experimental measurements, while allowing for optimization of the coincidence window to balance signal capture and accidental noise.

The heralding efficiency $\eta_{\mathrm{herald}}$ quantifies the probability of detecting one photon of a pair given the detection of its partner. For the signal or idler arm, the heralding efficiency is defined as:
\begin{equation}
    \eta_{\mathrm{herald,s/i}} = \frac{C_{\mathrm{true}}}{S_{\mathrm{measured,i/s}}},
\end{equation}
\noindent where $S_{\mathrm{measured},i/s}$ is the measured singles rate in the idler/signal arm. Note that the heralding efficiency is determined by the product of all possible sources of loss, from the generation of the pair of entangled photons, to their individual measurements. Thus, the overall heralding is influenced by: (i) $\eta_{\mathrm{static}}$ which refers to the static losses such as photon detection efficiency, losses in the EPS and coupling losses, and (ii) $\eta_{\mathrm{dynamic}}$ which represents losses arising from photon flux–dependent effects, such as detector dead time and afterpulsing. Therefore, the heralding $\eta$ can be expressed as:
\begin{equation}
    \eta = \eta_{\mathrm{static}} \eta_{\mathrm{dynamic}}.
\end{equation}
We characterized the heralding efficiencies of signal and idler photons as a function of pump power using two detection configurations driven in free-running operation (Fig.~\ref{fig:results}(c)) . Signal photons at 680~nm were detected with a Si-SPAD, while idler photons at 1550~nm were measured with either an InGaAs-SPAD (20 \% PDE, 20~$\upmu$s dead time) or a SNSPD (35 \% PDE, 25~ns dead time).

With the SNSPD, the signal heralding efficiency initially rises with pump power, from approximately 12 \% at 1~mW to a maximum of about 17.9 \% around 15--18~mW. This behavior reflects the increasing PGR, which enhances the heralding as true coincidence events begin to dominate over detector noise. At higher powers, a slight decrease in efficiency is observed, primarily due to saturation effects and the associated rise in accidental coincidences. The idler heralding efficiency exhibits an almost linear dependence on pump power, reaching about 33.8 \% near 4--5~mW and remaining stable (32--33 \%) at higher powers. The absence of saturation effects in this regime highlights the SNSPD’s short dead time and high maximum count rate, which allows it to follow the increasing photon flux.

In contrast, when using the InGaAs detector, the heralding performance is limited by detector-related effects. The long dead time and significant afterpulsing probability suppress the effective coincidence rate, resulting in a decrease of the signal heralding efficiency from approximately $2.1 \,\%$ to below $0.5 \,\%$. Although the idler efficiency still increases with pump power and saturates near $34 \,\%$, it is slightly distorted by detector dynamics. At low pump powers, dark counts contribute significantly to the singles rate, reducing the heralding efficiency. At higher pump powers, multipair generation leads to an increased rate of accidental coincidences, while afterpulsing further perturbs the measured statistics, particularly in the InGaAs case.

\subsection{Pair generation rate}

The pair generation rate $\text{PGR}$ can be estimated by \cite{Peng_2025}:
\begin{equation}
    \text{PGR} = \frac{S_{\mathrm{measured},s} \times S_{\mathrm{measured},i}}{C_{\mathrm{true}}}.
\end{equation}
From the measured singles on each channel and the coincidence counts, the PGR was estimated for each pump power. As shown in Fig.~\ref{fig:results}(d), the slope of the linear fit of PGR as a function of pump power provides the source brightness, i.e., the number of entangled photon pairs generated per second per mW of pump power. A slight difference in the slope was observed between the measurements performed with the InGaAs and SNSPD detectors, which can be attributed to their distinct afterpulsing characteristics. The spectral brightness $SB$ of the source was obtained by dividing the measured brightness by the corresponding spectral bandwidth, resulting in $SB_{\mathrm{SNSPD}} \approx 1870$ pairs/s/mW/GHz, and  $SB_{\mathrm{InGaAs}} \approx 1730$ pairs/s/mW/GHz.   

For heralding and brightness measurements, the coincidence time window was chosen such that $\Delta t \gg \sigma_{\mathrm{total}}$, such that the fraction $F \approx 1$, and thus $C_{\mathrm{true}} \approx C_{\mathrm{measured}} - C_{\mathrm{acc}}$ .

\subsection{Entanglement characterization}

\begin{figure}[h!]
\centering
\includegraphics[width=0.95\linewidth]{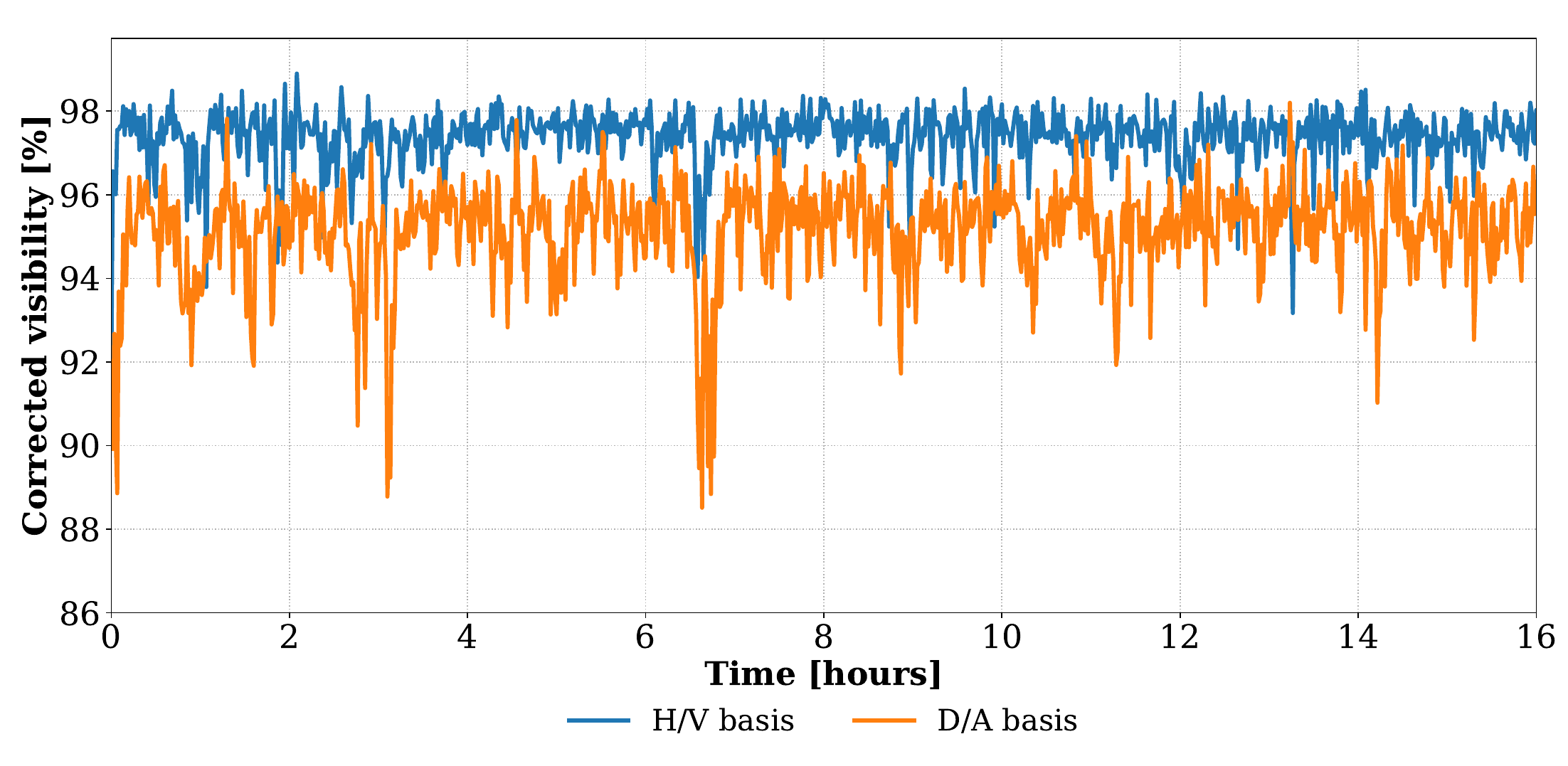}
\caption{\label{fig:visibility} The corrected visibilities in H/V and D/A basis over time.}
\end{figure}

The quality of polarization entanglement can be evaluated by measuring the polarization correlations in two mutually unbiased bases. From these measurements, the entanglement visibility \( {\cal V}_{\text{ab}} \) in the \(\{\text{a,b}\}\) basis is determined as \cite{dosan2022, Fazili:24}:
\begin{equation}
     {\cal V}_{\text{ab}} = \frac{C_{\mathrm{aa}} + C_{\mathrm{bb}} - C_{\mathrm{ab}} - C_{\mathrm{ba}}}{C_{\mathrm{aa}} + C_{\mathrm{bb}} + C_{\mathrm{ab}} + C_{\mathrm{ba}}},
\end{equation}
\noindent where \(C_{ij}\) denotes the coincidence counts measured when the signal photon is projected onto polarization state \(i\) and the idler photon onto state \(j\). We monitored the H/V and D/A visibilities over time, as shown in Fig.~\ref{fig:visibility}. The plots depict visibilities corrected for accidental coincidences. The entangled photon source exhibits excellent stability, yielding average visibilities of ${\cal V}_\text{HV} = (97.3 \pm 1.0)\,\%$ and ${\cal V}_\text{DA} = (94.9 \pm 1.6)\,\%$. The small variations observed are caused by polarization fluctuations in the optical fiber, which are actively compensated by the PolCo.

The fidelity $\cal F$ quantifies how close the experimentally prepared quantum state is to an ideal Bell state. A practical lower bound for the fidelity can be estimated from the averaged visibility measurements in two mutually unbiased bases: 
\begin{equation}
    {\cal F} \geq \frac{{\cal V}_{HV} + {{\cal V}_{DA}}}{2},
\end{equation}
Using the measured visibilities in the two mutually unbiased bases, we estimated the lower bound of the fidelity to the maximally entangled Bell state, ${\cal F} \geq (96.1\pm1.3)\,\%$. 
        
\section{Discussion}
The presented EPS provides a stable and compact architecture that bridges the visible and telecom domains. The heralding efficiencies were measured to be up to $\eta_{\text{signal}} \approx 18\,\%$ and $\eta_{\text{idler}} \approx 34\,\%$. These values reflect the intrinsic performance of the source under benchmark conditions, but the retrieved heralding efficiencies are also influenced by the detection scheme. Detector-specific parameters such as photon detection efficiency, dead time and afterpulsing introduce losses, thereby affecting the measured heralding. For this reason, SNSPD was combined with Si-SPAD to provide reference measurements under near-ideal conditions, while InGaAs-SPAD was used to assess the performance achievable in practical QKD systems.

Efficient fiber coupling in a crossed-crystal configuration presents challenges due to the difficulty of simultaneously overlapping the emission and collection waists emerging from both crystals. To achieve high heralding efficiencies, we chose a relatively large pump focus within the nonlinear crystals. This approach facilitates near single-spatial-mode biphoton emission and minimizes diffraction effects, thereby enhancing the spatial overlap between the emission and collection modes. However, increasing the pump focus size reduces the brightness of the source, introducing an inherent trade-off between achieving high heralding efficiency and maintaining high brightness. If sufficient pump power is available, a lower brightness may be acceptable as a compromise to obtain higher heralding.

The temporal evolution of the H/V and D/A visibilities demonstrates the stable operation of the source. The average measured visibilities of ${\cal V}_\text{HV} = (97.3 \pm 1.0)\,\%$ and ${\cal V}_\text{DA} = (94.9 \pm 1.6)\,\%$ correspond to a fidelity lower bound of ${\cal F} \geq (96.1\pm1.3)\,\%$. The residual deviation from unity fidelity arises primarily from partial distinguishability between the photon pairs generated in the two ppKTP crystals. Contributing factors include small spectral or spatial mismatches, unequal coupling efficiencies and residual temporal walk-off. Fine adjustment of the phase-compensation plate and careful alignment of the polarization analysis modules can further enhance the indistinguishability and improve the visibility.

The chosen crossed-crystal configuration provides a compact and robust architecture for generating polarization-entangled photon pairs without requiring interferometric stabilization. In this design, two orthogonally oriented ppKTP crystals share the same pump beam and emission axis, ensuring intrinsic phase stability and minimal spatial walk-off between the two SPDC processes. Only a static birefringent plate is needed to compensate for the temporal and phase delay between the orthogonal contributions. The collinear geometry simplifies alignment and enables efficient fiber coupling with high spatial mode overlap. This approach is particularly well suited for highly nondegenerate operation, where interferometric schemes such as Sagnac or Mach–Zehnder configurations become challenging to align and require complex optics to maintain compatibility across multiple wavelengths. Consequently, the crossed-crystal design offers a practical balance between simplicity, stability, and performance, making it an ideal choice for wavelength-hybrid entangled photon sources.

Future developments could further enhance the source performance. Substituting ppKTP with periodically poled lithium niobate (ppLN) may provide a narrower spectral profile and increased brightness owing to its higher nonlinear coefficient. However, ppLN is considerably more temperature-sensitive, and current poling techniques restrict the achievable crystal aperture, which can limit coupling efficiency and mechanical robustness. Another possible development involves adopting a single-crystal geometry, such as a Sagnac interferometer, which can improve spatial mode overlap and fiber coupling while enabling longer nonlinear interaction lengths. However, this configuration is more difficult to align and requires additional, complex optics to ensure compatibility with highly nondegenerate operation. Despite these trade-offs, both strategies offer promising routes toward more efficient and stable entangled photon sources for long-distance QKD and quantum network applications.

\section{Conclusions}

We have demonstrated a highly nondegenerate EPS based on a 473 nm pump generating 680/1550 nm photon pairs via SPDC in a crossed-crystal configuration. The source exhibits an SPDC bandwidth of 301 GHz (2.4 nm), in excellent agreement with the 300 GHz theoretical prediction, and achieves spectral brightness up to ~ $1.9~\times~10^3$~pairs~s$^{-1}$~mW$^{-1}$~GHz$^{-1}$. Heralding efficiencies reach $\eta_{\text{signal}} \approx 18\,\%$ and $\eta_{\text{idler}} \approx 34\,\%$, and the measured visibilities of ${\cal V}_\text{HV} = (97.3 \pm 1.0)\,\%$ and ${\cal V}_\text{DA} = (94.9 \pm 1.6)\,\%$ correspond to a fidelity lower bound of ${\cal F} \geq (96.1\pm1.3)\,\%$. By simultaneously making use of the peak detection efficiency of Si-SPADs around 680 nm and the low fiber attenuation in the telecom C-band at 1550 nm, the presented wavelength-hybrid EPS constitutes a practical, high-performance building block for future fiber-based QKD links and emerging quantum network architectures.

\section*{Acknowledgements}

The authors acknowledge support from a grant of The Federal Ministry of Research, Technology and Space (BMFTR), project number 16KIS2163K (SQuIRRL).

\bibliographystyle{unsrt}
\bibliography{References}

\end{document}